\documentclass[preprint,12pt]{elsarticle}

\usepackage{amsmath,amssymb,amsfonts}
\usepackage{graphicx}
\usepackage{booktabs}
\usepackage{multirow}
\usepackage{siunitx}
\usepackage{algorithm}
\usepackage{algpseudocode}
\usepackage{float}
\usepackage{url}
\usepackage{hyperref}
\usepackage{enumitem}
\usepackage{mathtools}
\usepackage{tikz}
\usetikzlibrary{arrows.meta,positioning}
\journal{Applied Energy}

\begin{document}

\begin{frontmatter}

\title{Operational Value of Multi-Horizon Load Forecasts for Energy Management in a Hydrogen-Enabled Community Microgrid}

\author[inst1]{Mohamed Atef}
\ead{Muhamed.Atef1408@gmail.com}

\author[inst1]{Sanath Alahakoon}
\author[inst2]{Umme Mumtahina}
\author[inst2]{Peter Wolfs}
\author[inst3]{Tamer Khatib}
\author[inst4]{Moslem Uddin}

\address[inst1]{School of Engineering and Technology, Central Queensland University, Gladstone, QLD 4680, Australia}
\address[inst2]{School of Engineering and Technology, Central Queensland University, Rockhampton, QLD 4680, Australia}
\address[inst3]{Energy Engineering \& Environment Dept., An-Najah National University, Nablus, West Bank 97300, Palestine}
\address[inst4]{School of Engineering \& Technology, The University of New South Wales, Canberra, ACT 2610, Australia}

\begin{abstract}

Hydrogen-enabled community microgrids can improve renewable energy utilization and local resilience, but their operation is complicated by uncertain residential demand, variable renewable generation, dynamic electricity prices, and the coupled dynamics of battery and hydrogen storage. This paper evaluates the operational value of multi-horizon community-load forecasts when they are added to a previously developed proximal policy optimization (PPO) energy management system. The framework is studied for a 1000-household residential microgrid in Rockhampton, Australia. Forecast accuracy is mixed: the 1-h horizon achieves an RMSE of \SI{239.32}{\kilo\watt} and $R^2=0.201$, whereas the 6-h and 12-h horizons produce negative $R^2$ values; the 24-h forecast achieves an RMSE of \SI{249.79}{\kilo\watt}, MAPE of 62.52\%, and $R^2=0.126$. In the reported single-seed PPO experiment, forecast-enriched control reaches a final reward 8.3\% above the non-predictive controller. Annual savings rise from A\$2,439.86 with non-predictive PPO to A\$2,765.83 with forecast-enriched PPO, an incremental gain of A\$325.97 or 13.4\% relative to the non-predictive savings, while grid imports fall to \SI{58147.49}{\kilo\watt\hour}. These results constitute proof-of-concept evidence rather than a multi-seed validation.

\end{abstract}


\end{frontmatter}


\section{Introduction}

\subsection{Background}

Community microgrids are increasingly being adopted as decentralized energy platforms that improve local reliability, increase renewable energy utilization, and support the transition toward low-carbon electricity systems \cite{uddin2023microgrids,chartier2022microgrid}. Community-scale studies further emphasize that local distributed energy resource (DER) coordination can support resilient and low-carbon operation when distributed generation, storage, controllable demand, and grid exchange are managed together \cite{trivedi2022community,atef2026energy}. Recent residential microgrid research shows a shift from conventional photovoltaic (PV)--battery configurations toward multi-energy platforms that include PV, wind generation, battery energy storage systems (BESSs), demand response, and hydrogen-based storage pathways \cite{atef2026review,sadeghian2024energy}.

Battery storage is effective for short-term balancing, peak shaving, and renewable self-consumption. However, battery-only systems can be limited when renewable shortfalls persist for several hours or days. Hydrogen technologies provide a complementary long-duration storage route in which surplus renewable electricity is converted to hydrogen through an electrolyzer, stored in a hydrogen tank, and later reconverted to electricity by a fuel cell \cite{modu2023systematic,elkhatib2024green}. Techno-economic and sustainability studies also show that hydrogen pathways can improve renewable hosting and long-duration autonomy when they are coordinated with electrical storage \cite{das2024sustainable,enaloui2025techno}. This battery--hydrogen architecture is particularly suitable for community microgrids because batteries can respond rapidly to short-term power fluctuations, while hydrogen storage can support resilience and autonomy over longer time scales.

The operational value of such systems depends strongly on how well the energy management system (EMS) anticipates uncertainty. Solar irradiance, wind speed, residential demand, and electricity market prices vary continuously and can lead to inefficient storage dispatch if the EMS relies only on current measurements. The integrated Australian microgrid design framework established the PV--wind--battery--hydrogen system foundation considered here \cite{atef2026integrated}. A subsequent PPO-based preprint formulated learning-based dispatch for the same 1000-household community \cite{atef2026deepreinforcementlearningbasedenergy}. The present paper extends that controller by adding a forecasting layer, allowing the policy to use expected community demand rather than reacting only to instantaneous observations.

\subsection{Motivation}

Conventional EMS approaches, including rule-based control, deterministic scheduling, and metaheuristic optimization, remain useful because they are interpretable and can be implemented with modest computational requirements. Nevertheless, fixed dispatch rules cannot fully adapt to rapid changes in renewable production, household demand, electricity prices, or storage availability \cite{Houben2023,Majeed2023}. Recent EMS studies also note that deterministic and heuristic dispatch strategies can become sensitive to forecast assumptions and operating-condition changes \cite{Niknami2024,Tang2024}. Optimization-based methods can improve dispatch by solving cost, emission, and reliability objectives, but they may require repeated computation whenever operating conditions change \cite{Asif2024,Bhuvaneshwarri2025}. Reliability-oriented hybrid-system planning has also optimized solar-PV and gas-turbine capacity using the loss-of-load probability index \cite{atef2020optimization}. A data-driven optimization framework has combined demand response with uncertain power generation to improve the efficiency and reliability of microgrid energy management \cite{atef2024datadriven}. These limitations become more important in hydrogen-enabled community microgrids, where the EMS must simultaneously schedule short-term battery actions, long-duration hydrogen production and use, fuel-cell dispatch, and grid import/export decisions.

Forecasting provides future awareness that can improve sequential storage and grid decisions. Multi-step community-load forecasts can help the EMS preserve battery capacity before demand peaks, avoid unnecessary charging before low-demand periods, and shift grid transactions toward economically favorable intervals. The present study treats the forecast series as externally generated information and evaluates their operational value within the PPO EMS in \cite{atef2026deepreinforcementlearningbasedenergy}; it does not claim to propose or validate the underlying forecasting architecture.

\subsection{Research Gap}

Recent EMS literature highlights the importance of uncertainty-aware control and forecasting for renewable-rich microgrids \cite{cabrera2022review,sharma2022critical}. Other reviews emphasize data-driven scheduling, resilience, and scalable EMS design as key requirements for future microgrids \cite{allwyn2023comprehensive,bilal2024review}. Recent work also points to persistent gaps in translating advanced EMS methods into robust, sector-specific operation under uncertainty \cite{mannan2024recent}. Existing work has separately advanced renewable/load forecasting, hydrogen-enabled microgrid planning, and DRL-based dispatch. However, many forecasting studies evaluate prediction accuracy without demonstrating how forecast outputs change battery and hydrogen operation, while many DRL EMS studies learn dispatch policies from current states without explicitly embedding multi-step forecasts into the control state. Similarly, hydrogen microgrid studies often focus on sizing, techno-economic assessment, or deterministic scheduling, leaving a need for integrated predictive control frameworks that couple advanced forecasting with adaptive learning-based dispatch.

The previous planning framework established the PV--wind--battery--hydrogen community configuration and techno-economic foundation \cite{atef2026integrated}. The subsequent PPO study demonstrated learning-based dispatch but did not provide explicit future information to the controller \cite{atef2026deepreinforcementlearningbasedenergy}. The remaining research gap is therefore the evaluation of a forecast-enriched PPO framework in which multi-horizon predictions are embedded into the EMS state, together with a quantitative assessment of whether imperfect forecasts improve learning and operation.

\subsection{Contributions}

The main contributions of this paper are summarized as follows:

\begin{itemize}[leftmargin=*]
    \item A multi-horizon forecasting layer is evaluated for 1, 6, 12, 24, 48, and 72-h community-load prediction using RMSE, MAE, MAPE, nRMSE, correlation, and $R^2$.
    \item A predictive EMS is developed by extending the PPO-based hydrogen-enabled microgrid controller in \cite{atef2026deepreinforcementlearningbasedenergy} with forecast-enriched state information.
    \item Forecast-enriched PPO, non-predictive PPO, perfect-forecast operation, and no-battery operation are compared to isolate the operational value of forecast information.
    \item The framework is evaluated using a 1000-household Australian residential microgrid based on the earlier planning studies \cite{atef2026integrated,atef2025techno}.
\end{itemize}

Table~\ref{tab:novelty_comparison} distinguishes the present contribution from the earlier planning and PPO studies. System sizing and the battery--hydrogen plant model are inherited, while the new contribution is the explicit evaluation of multi-horizon forecast information within PPO control.

\begin{table}[H]
\centering
\caption{Relationship between the present study and the preceding planning and PPO studies.}
\label{tab:novelty_comparison}
\small
\setlength{\tabcolsep}{3pt}
\begin{tabular*}{\textwidth}{@{\extracolsep{\fill}}lccc@{}}
\toprule
Feature & Planning study & PPO study & Present study \\
\midrule
System sizing & \checkmark & Inherited & Inherited \\
Battery--hydrogen model & \checkmark & \checkmark & \checkmark \\
PPO control & -- & \checkmark & \checkmark \\
Multi-horizon load forecasts & -- & -- & \checkmark \\
Forecast-value analysis & -- & -- & \checkmark \\
Perfect-forecast benchmark & -- & -- & \checkmark \\
Forecast-specific resilience benefit & -- & -- & Not demonstrated \\
\bottomrule
\end{tabular*}
\end{table}

\subsection{Paper Organization}

The remainder of this paper is organized as follows. Section~\ref{sec:SystemDescription} presents the architecture and mathematical models of the hydrogen-enabled community microgrid. Section~\ref{sec:ForecastingFramework} introduces the multi-horizon load-forecasting framework, input features, model architecture, training process, and evaluation metrics. Section~\ref{sec:PredictiveEMS} describes the forecast-enriched EMS and its integration with the PPO controller. Section~\ref{sec:CaseStudy} presents the Australian community microgrid case study, datasets, simulation settings, and benchmark methods. Section~\ref{sec:Results} analyzes forecasting accuracy and the resulting impacts on microgrid operation. Section~\ref{sec:Discussion} discusses key findings and practical implications. Finally, Section~\ref{sec:Conclusion} concludes the paper and outlines future research directions.

\section{System Description}
\label{sec:SystemDescription}

This section describes the hydrogen-enabled community microgrid used to evaluate the forecast-enriched EMS\@. The model follows the same physical system considered in \cite{atef2026integrated}, but it is prepared here for forecast-enriched operation by explicitly defining the renewable generation, demand, battery, hydrogen, and grid-balance equations used by the simulation environment. The community scale used in this paper is fixed at $N_h=1000$ households, and all aggregated demand and DER scheduling variables are interpreted at this community level rather than the smaller 100-household model.

\subsection{Microgrid Architecture}

The studied system supplies an aggregated residential community through a hybrid renewable and storage infrastructure. PV arrays and wind turbines are the primary local generation resources. A BESS provides short-duration electrical storage, absorbs renewable surplus during high-generation intervals, and supports the load during short-term renewable deficits. The hydrogen subsystem consists of an electrolyzer, a hydrogen storage tank, and a fuel cell. When local renewable generation exceeds community demand and battery-charging requirements, the electrolyzer converts part of the surplus electricity into hydrogen. When renewable production and battery discharge are insufficient, the fuel cell can reconvert stored hydrogen into electricity.

The microgrid is also connected to the utility grid. Grid import covers residual demand after local renewable generation and storage dispatch, while grid export allows unused renewable electricity to be transferred outside the community boundary when local storage and conversion options are saturated or economically unattractive. The EMS coordinates PV and wind utilization, battery charging/discharging, electrolyzer power, fuel-cell output, hydrogen storage, and grid transactions. In the predictive framework, this coordination is informed by current measurements and externally generated multi-horizon community-load forecasts.

\begin{figure}[H]
\centering
\begin{tikzpicture}[
x=1mm,
y=1mm,
font=\small,
box/.style={draw, rounded corners, minimum width=25mm, minimum height=8.5mm, align=center, fill=gray!8},
source/.style={box, fill=green!10},
storage/.style={box, fill=blue!10},
control/.style={box, fill=orange!12},
load/.style={box, fill=red!8},
flow/.style={-{Latex[length=2.5mm]}, thick},
biflow/.style={{Latex[length=2.5mm]}-{Latex[length=2.5mm]}, thick},
signal/.style={-{Latex[length=2.5mm]}, dashed, thick}
]
\node[source] (pv) at (-54,15) {PV array};
\node[source] (wind) at (-54,5) {Wind turbine};
\node[box] (grid) at (-54,-5) {Utility grid};
\node[source] (diesel) at (-54,-15) {Diesel backup};
\node[box, minimum width=30mm, minimum height=14mm] (bus) at (0,0) {Community\\AC bus};
\node[load, minimum width=29mm] (load) at (54,0) {1000-household\\community load};
\node[storage] (battery) at (-12,-23) {Battery\\SOC};
\node[storage] (ely) at (18,-23) {Electrolyzer};
\node[storage] (tank) at (50,-23) {Hydrogen\\tank};
\node[storage] (fc) at (50,-11) {Fuel cell};
\node[control, minimum width=31mm] (ppo) at (0,28) {PPO energy\\management policy};
\node[control, minimum width=27mm] (forecast) at (-40,28) {Multi-horizon\\load forecasts};

\draw[flow] (pv.east) -- ([yshift=6mm]bus.west);
\draw[flow] (wind.east) -- ([yshift=2mm]bus.west);
\draw[biflow] (grid.east) -- ([yshift=-2mm]bus.west);
\draw[flow] (diesel.east) -- ([yshift=-6mm]bus.west);
\draw[flow] (bus.east) -- (load.west);
\draw[biflow] ([xshift=-5mm]bus.south) -- (battery.north);
\draw[flow] ([xshift=5mm]bus.south) -- (ely.north);
\draw[flow] (ely.east) -- node[above]{H$_2$} (tank.west);
\draw[flow] (tank.north) -- (fc.south);
\draw[flow] (fc.west) -- ++(-6,0) |- ([yshift=-3.5mm]bus.east);
\draw[signal] (forecast.east) -- (ppo.west);
\draw[signal] (ppo.south) -- node[right, font=\scriptsize]{dispatch} (bus.north);
\end{tikzpicture}
\caption{
Hydrogen-enabled community microgrid architecture showing both physical energy flows (solid arrows) and communication/control signal flows (dashed arrows). The multi-horizon forecasting module provides future load information to the PPO-based energy management system, which generates supervisory control commands for the battery energy storage system, hydrogen subsystem, diesel backup, and grid interaction.}
\label{fig:system_architecture}
\end{figure}
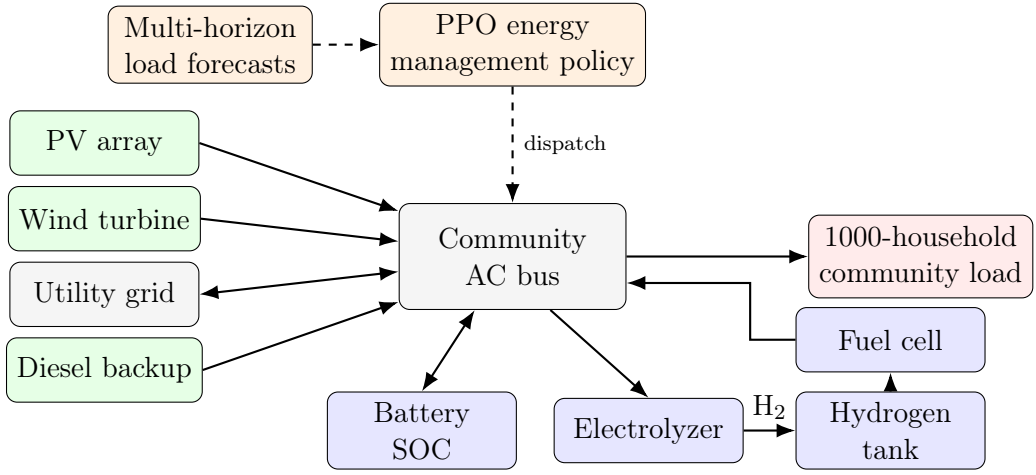

Figure~\ref{fig:system_architecture} summarizes the physical energy paths and information flow. Solid arrows denote electrical or hydrogen-energy transfer, while dashed arrows denote forecast and PPO control signals.

\subsection{Mathematical Models}

The microgrid is represented in discrete time with interval length $\Delta t$. At each time step $t$, the community electrical demand is denoted by $P_{L,t}$. If individual household demand profiles $p_{h,t}$ are available, the aggregate community load is
\begin{equation}
P_{L,t}=\sum_{h=1}^{N_h}p_{h,t},
\label{eq:community_load}
\end{equation}
where $N_h$ is the number of households. If a representative normalized household profile is used for all customers, Eq.~\eqref{eq:community_load} becomes $P_{L,t}=N_h\,p_{h,t}$. Alternatively, when a base profile $P_{b,t}$ is available for $N_b$ households, the scaled load is expressed as
\begin{equation}
P_{L,t}=\alpha_L P_{b,t},\qquad \alpha_L=\frac{N_h}{N_b},
\label{eq:scaled_load}
\end{equation}
where $\alpha_L$ is the demand-scaling factor.

\subsubsection{Photovoltaic Generation Model}

The PV output is modeled as a function of installed capacity, solar irradiance, derating losses, and cell-temperature effects \cite{zia2022energy,Witharama2024}. The generated PV power is calculated as
\begin{equation}
P_{PV,t}=P_{PV}^{rated} f_{PV} G_t
\left[1+\gamma_{PV}\left(T_{c,t}-T_c^{ref}\right)\right],
\label{eq:pv_model}
\end{equation}
where $P_{PV,t}$ is the PV power output, $P_{PV}^{rated}$ is the installed PV capacity, $f_{PV}$ is the derating factor, $G_t=I_t/I_{ref}$ is the normalized irradiance, $I_t$ is the measured irradiance, $I_{ref}$ is the reference irradiance, $\gamma_{PV}$ is the PV temperature coefficient, $T_{c,t}$ is the PV cell temperature, and $T_c^{ref}$ is the reference cell temperature. The cell temperature is estimated from ambient temperature and irradiance as
\begin{equation}
T_{c,t}=T_{amb,t}+\left(\frac{T_{NOCT}-T_{amb}^{NOCT}}{I_{NOCT}}\right)I_t,
\label{eq:pv_temperature}
\end{equation}
where $T_{amb,t}$ is the ambient temperature, $T_{NOCT}$ is the nominal operating cell temperature, $T_{amb}^{NOCT}=\SI{20}{\celsius}$, and $I_{NOCT}=\SI{800}{\watt\per\square\meter}$.

\subsubsection{Wind Turbine Model}

The wind turbine output is represented using the aerodynamic power relationship and constrained by the practical turbine operating curve \cite{zia2022energy}. The available wind power is
\begin{equation}
P_{WT,t}=\frac{1}{2}\rho C_p A_s v_t^3,
\label{eq:wind_model}
\end{equation}
where $P_{WT,t}$ is the wind turbine power, $\rho$ is air density, $C_p$ is the power coefficient, $A_s$ is the swept rotor area, and $v_t$ is the hub-height wind speed. In implementation, Eq.~\eqref{eq:wind_model} is bounded by cut-in, rated, and cut-out wind-speed limits so that the turbine produces zero power below cut-in and above cut-out speeds, follows the cubic region below rated speed, and is capped at rated power in the rated operating region.

\subsubsection{Battery Energy Storage Model}

The BESS state of charge (SOC) is updated according to the charging and discharging powers selected by the EMS \cite{yang2022modelling}. The SOC transition is
\begin{equation}
SOC_{t+1}=SOC_t+\frac{\Delta t}{E_B^{rated}}
\left(\eta_{ch}P_{B,t}^{ch}-\frac{P_{B,t}^{dis}}{\eta_{dis}}\right),
\label{eq:battery_soc}
\end{equation}
where $SOC_t$ is the battery state of charge, $E_B^{rated}$ is the rated battery energy capacity, $P_{B,t}^{ch}$ and $P_{B,t}^{dis}$ are the battery charging and discharging powers, and $\eta_{ch}$ and $\eta_{dis}$ are the charging and discharging efficiencies. The SOC is constrained by
\begin{equation}
\underline{SOC}\leq SOC_t\leq \overline{SOC},
\label{eq:battery_soc_limits}
\end{equation}
where $\underline{SOC}$ and $\overline{SOC}$ define the minimum and maximum allowable SOC values. These limits prevent deep discharge and overcharge, thereby preserving feasible and safe battery operation.

The signed PPO battery action $a_{B,t}\in[-1,1]$ is mapped directly to mutually exclusive charging and discharging powers:
\begin{align}
P_{B,t}^{ch}&=\max(-a_{B,t},0)\overline{P}_{B}^{ch},
&
P_{B,t}^{dis}&=\max(a_{B,t},0)\overline{P}_{B}^{dis},
\label{eq:battery_power_limits}
\end{align}
where $\overline{P}_{B}^{ch}$ and $\overline{P}_{B}^{dis}$ are the charging and discharging limits. This deterministic mapping prevents simultaneous charging and discharging without introducing a mixed-integer decision variable.

\subsubsection{Hydrogen Subsystem Model}

The hydrogen subsystem is formulated on a mass basis to remain consistent with the tank capacity and flow limits reported in kilograms. Let $L_H$ denote the hydrogen lower heating value in \si{\kilo\watt\hour\per\kilogram}. Electrolyzer production and fuel-cell consumption are related to electrical power by
\begin{align}
\dot m_{H,t}^{prod} &= \frac{\eta_{ely}P_{ely,t}}{L_H},
&
\dot m_{H,t}^{cons} &= \frac{P_{FC,t}}{\eta_{fc}L_H},
\label{eq:hydrogen_conversion}
\end{align}
where $\dot m_{H,t}^{prod}$ and $\dot m_{H,t}^{cons}$ are hydrogen production and consumption rates in \si{\kilogram\per\hour}, and $\eta_{ely}$ and $\eta_{fc}$ are electrical-to-chemical and chemical-to-electrical efficiencies, respectively. The tank state evolves as
\begin{equation}
m_{H,t+1}=m_{H,t}+\left(\dot m_{H,t}^{prod}-\dot m_{H,t}^{cons}\right)\Delta t,
\label{eq:hydrogen_storage}
\end{equation}
subject to
\begin{align}
\underline{m}_H &\leq m_{H,t}\leq \overline{m}_H, \\
0&\leq \dot m_{H,t}^{prod}\leq \overline{\dot m}_{H}^{prod},
&
0&\leq \dot m_{H,t}^{cons}\leq \overline{\dot m}_{H}^{cons},
\label{eq:hydrogen_limits}
\end{align}
where $\underline{m}_H$ and $\overline{m}_H$ are the tank limits. The signed hydrogen action $a_{H,t}\in[-1,1]$ maps directly to electrolyzer and fuel-cell power:
\begin{align}
P_{ely,t}&=\max(a_{H,t},0)\overline{P}_{ely},
&
P_{FC,t}&=\max(-a_{H,t},0)\overline{P}_{FC}.
\label{eq:hydrogen_exclusivity}
\end{align}

\subsubsection{Grid and Backup-Generator Constraints}

Let $R_t$ denote residual demand before grid exchange:
\begin{align}
R_t={}&P_{L,t}+P_{B,t}^{ch}+P_{ely,t}
-P_{PV,t}-P_{WT,t}-P_{FC,t}-P_{B,t}^{dis}-P_{DG,t}.
\label{eq:residual_demand}
\end{align}
Grid import and export are then calculated deterministically as
\begin{align}
P_{grid,t}^{buy}&=\min\!\left(\max(R_t,0),\overline{P}_{grid}^{buy}\right),\\
P_{grid,t}^{sell}&=\min\!\left(\max(-R_t,0),\overline{P}_{grid}^{sell}\right),
\label{eq:grid_limits}
\end{align}
The backup diesel generator is represented by
\begin{align}
0&\leq P_{DG,t}\leq o_t\overline{P}_{DG},
&
-R_{DG}^{down}&\leq P_{DG,t}-P_{DG,t-1}\leq R_{DG}^{up},
\label{eq:diesel_limits}
\end{align}
where $o_t=1$ when backup generation is enabled, $\overline{P}_{DG}$ is rated power, and $R_{DG}^{up}$ and $R_{DG}^{down}$ are ramp limits. If fuel use and emissions are evaluated, they are calculated from the generator efficiency and fuel properties rather than from electrical output alone.

\subsubsection{Power Balance Equation}

At every time step, the EMS must satisfy the electrical power balance between local generation, storage actions, hydrogen conversion, grid exchange, and community demand. The balance is written as
\begin{align}
&P_{PV,t}+P_{WT,t}+P_{FC,t}+P_{B,t}^{dis}+P_{DG,t}
+P_{grid,t}^{buy}+P_t^{unmet} \notag\\
&\qquad-P_{L,t}-P_{B,t}^{ch}-P_{ely,t}-P_{grid,t}^{sell}-P_t^{curt}=0,
\label{eq:power_balance}
\end{align}
where $P_t^{unmet}=\max(R_t-P_{grid,t}^{buy},0)$ records unserved demand and $P_t^{curt}=\max(-R_t-P_{grid,t}^{sell},0)$ records curtailed surplus. Each normalized PPO action is clipped to its component range and reconciled through Eq.~\eqref{eq:power_balance}. This feasibility mapping prevents the policy from directly applying an inadmissible physical command.


\section{Multi-Horizon Load Forecast Information}
\label{sec:ForecastingFramework}

This section defines the externally generated forecast information evaluated by the predictive EMS\@. Community-load forecasts are available at 1, 6, 12, 24, 48, and 72-h horizons. The paper evaluates whether these forecasts improve PPO operation; forecasting-model development and architectural comparison are outside its scope. Data-driven renewable-system modeling and AI-enabled EMS studies provide the broader forecasting context \cite{atef2019utilization,javed2025sustainable,tabaku2025utilizing}, while hybrid LSTM--XGBoost and GRU-attention methods illustrate alternative nonlinear time-series approaches \cite{dakheel2025optimizing,boucetta2025wind}.

\subsection{Forecasting Problem Formulation}

At each decision instant $t$, the external forecasting process generates a community-load estimate for each evaluated horizon $k$:
\begin{equation}
\widehat{P}_{L,t+k}=\mathcal{F}_{k}\!\left(\mathcal{I}_t\right),
\label{eq:forecast_mapping}
\end{equation}
where $\mathcal{I}_t$ denotes information available at time $t$ and $k\in\mathcal{K}$, with
\begin{equation}
\mathcal{K}=\{1,6,12,24,48,72\}\ \mathrm{h}.
\label{eq:forecast_targets}
\end{equation}
Here, $\widehat{P}_{L,t+k}$ is the predicted aggregated load of the 1000-household community. Equation~\eqref{eq:forecast_mapping} is intentionally architecture-neutral because the present contribution concerns forecast value in control rather than forecasting-model novelty.

\subsection{Data Collection and Feature Engineering}

The underlying microgrid dataset contains 8,760 hourly observations of community demand, available PV and wind power, grid purchase price, and export price. The forecasting inputs are aligned to hourly resolution and normalized using training-set statistics.

The feature design follows established data-driven EMS and uncertainty-aware scheduling practices \cite{javed2025sustainable,quispe2025energy,ramkumar2025optimal}.

\subsection{Input Features and Forecast Horizon}

The general input vector available to the forecasting workflow is
\begin{equation}
\mathbf{x}_t=\left[I_t,T_{amb,t},v_t,P_{PV,t},P_{WT,t},P_{L,t},\lambda_t,\phi_t\right],
\label{eq:forecast_input_vector}
\end{equation}
where $I_t$ is irradiance, $T_{amb,t}$ is ambient temperature, $v_t$ is wind speed, $\lambda_t$ is the electricity price, and $\phi_t$ represents temporal indicators. The evaluated horizons are $H\in\{1,6,12,24,48,72\}$ h. Short horizons support near-real-time battery scheduling, whereas day-ahead and multi-day horizons can inform energy-reserve planning. The empirical results show that forecast quality is strongly horizon-dependent rather than monotonically decreasing with horizon.

\subsection{Forecasting Performance Metrics}

Forecasting performance is evaluated using RMSE, MAE, MAPE, normalized RMSE (nRMSE), and the coefficient of determination $R^2$. These metrics are defined as
\begin{align}
\mathrm{RMSE} &= \sqrt{\frac{1}{N}\sum_{i=1}^{N}\left(y_i-\widehat{y}_i\right)^2},
&
\mathrm{MAE} &= \frac{1}{N}\sum_{i=1}^{N}\left|y_i-\widehat{y}_i\right|, \\
\mathrm{MAPE} &= \frac{100}{N}\sum_{i=1}^{N}\left|\frac{y_i-\widehat{y}_i}{y_i}\right|,
&
R^2 &= 1-\frac{\sum_{i=1}^{N}(y_i-\widehat{y}_i)^2}{\sum_{i=1}^{N}(y_i-\bar{y})^2}.
\label{eq:forecast_metrics}
\end{align}
RMSE emphasizes large errors, MAE measures average absolute deviation, MAPE provides a percentage-based interpretation, nRMSE normalizes RMSE by the data scale, and $R^2$ indicates the proportion of variance explained by the model. Correlation is also reported to distinguish pattern tracking from amplitude accuracy. The available benchmark set comprises linear, ridge, lasso, autoregressive (AR), and time-of-day (TOD) average models.

\section{Predictive Energy Management Framework}
\label{sec:PredictiveEMS}

This section presents the predictive EMS that supplies the load forecast to the PPO controller developed in the preceding DRL study \cite{atef2026deepreinforcementlearningbasedenergy}. PPO is retained so that the value of forecasting can be isolated without simultaneously changing the control algorithm. The proposed framework integrates three complementary components: MHLF, a DRL-based EMS, and the PPO algorithm. The MHLF module operates independently of the controller and generates future community load predictions at multiple forecasting horizons. These forecasts are incorporated into the EMS by augmenting the state representation available to the DRL agent. Within the EMS, PPO serves as the DRL algorithm responsible for learning the optimal control policy from interactions with the hydrogen-enabled community microgrid. Thus, MHLF enhances the information available to the controller, DRL provides the learning-based decision-making framework, and PPO is the optimization algorithm used to determine battery, hydrogen, diesel, and grid dispatch actions.
DRL is appropriate because it can learn sequential control policies under nonlinear dynamics and uncertainty. Recent applications have demonstrated DRL for PV-aware railway control, smart-home demand response, safe building operation, and coupled port-energy scheduling \cite{xu2026drl,abishu2025multi}. Constraint-aware and degradation-aware studies further motivate careful reward design for hybrid energy systems \cite{wang2025safe,liu2025novel}.

\subsection{Predictive MDP Formulation}

The EMS is formulated as a forecast-enriched Markov Decision Process (MDP) defined by the tuple $(\mathcal{S},\mathcal{A},\mathcal{P},r,\gamma)$. The environment is the hydrogen-enabled community microgrid, $\mathcal{S}$ is the state space, $\mathcal{A}$ is the continuous action space, $\mathcal{P}$ is the transition function governed by Section~\ref{sec:SystemDescription}, $r$ is the hourly reward, and $\gamma$ is the discount factor. The PPO policy $\pi_{\theta}$ maximizes
\begin{equation}
J(\theta)=\mathbb{E}_{\pi_{\theta}}\left[\sum_{t=0}^{T-1}\gamma^t r_t\right].
\label{eq:ppo_return}
\end{equation}
At each hour, the agent observes current operating conditions and the forecast vector, selects battery, diesel, and hydrogen commands, and receives the next state after residual grid exchange and storage updates are calculated.

The MDP formulation follows the general structure used in recent DRL energy management studies, where uncertain PV generation and load disturbance are represented through state transitions and reward feedback \cite{xu2026drl}. For demand-side and transportation-coupled applications, MDP-based DRL has also been used to represent user behavior, comfort, logistics, and energy scheduling interactions \cite{abishu2025multi,song2025two}. In contrast to non-predictive DRL, the proposed MDP augments the state with external multi-horizon load forecasts, allowing the policy to anticipate future demand changes.

\subsection{Forecast-Enriched State Space}

The base PPO observation contains battery SOC, normalized hydrogen level, time features, normalized PV, wind, load, purchase and export prices, and the grid-availability indicator. The forecast-enriched state appends the multi-horizon load forecast:
\begin{equation}
\begin{aligned}
\mathbf{s}^{F}_t=\big[&SOC_t,H_t,\tau_t,\sin(2\pi\tau_t),\cos(2\pi\tau_t),\\
&\widetilde{P}_{PV,t},\widetilde{P}_{WT,t},\widetilde{P}_{L,t},
\widetilde{\pi}^{buy}_t,\widetilde{\pi}^{sell}_t,G_t,
\widetilde{\widehat{\mathbf{P}}}_{L,t+1:t+H}\big]^{\top}.
\end{aligned}
\label{eq:forecast_state}
\end{equation}
Here, $H_t$ is the normalized hydrogen storage level, $\tau_t$ is normalized hour of day, $G_t$ is the outage indicator, and tildes denote normalized values. The non-predictive controller receives the same base state without $\widetilde{\widehat{\mathbf{P}}}_{L,t+1:t+H}$, allowing a controlled comparison of forecast value.

\subsection{Action Space}

{\raggedright The PPO agent retains the three continuous actions of the base manuscript:\par}

\begin{equation}
\mathbf{a}_t=\left[a_{B,t},a_{DG,t},a_{H,t}\right]^{\top},\qquad
a_{B,t},a_{H,t}\in[-1,1],\quad a_{DG,t}\in[0,1].
\label{eq:action_space}
\end{equation}
The battery command is scaled by \SI{530}{\kilo\watt}; positive values discharge and negative values charge. The diesel command is scaled by \SI{180}{\kilo\watt} and is enabled during outages. Positive hydrogen actions operate the electrolyzer, while negative values dispatch the fuel cell. Grid import or export is calculated from the residual power balance rather than selected directly by the policy.

\subsection{Reward Function}

Actual grid transactions and component-use charges are separated from reward-shaping penalties. For an interval of length $\Delta t$, the operating-cost term is
\begin{equation}
\begin{aligned}
C_t^{op}={}&\left(\pi_t^{buy}P_{grid,t}^{buy}-\pi_t^{sell}P_{grid,t}^{sell}
+c_B\left(P_{B,t}^{ch}+P_{B,t}^{dis}\right)\right)\Delta t\\
&+c_{DG}P_{DG,t}\Delta t+c_H^{prod}\dot m_{H,t}^{prod}\Delta t
+c_H^{cons}\dot m_{H,t}^{cons}\Delta t,
\end{aligned}
\label{eq:operating_cost}
\end{equation}
where $\pi_t^{buy}$ and $\pi_t^{sell}$ are in A\$/\si{\kilo\watt\hour}, $c_B$ and $c_{DG}$ are in A\$/\si{\kilo\watt\hour}, and $c_H^{prod}$ and $c_H^{cons}$ are in A\$/\si{\kilogram}. Export revenue enters with a negative sign. The training reward is
\begin{equation}
r_t=-C_t^{op}-\lambda_{unmet}P_t^{unmet}\Delta t
-\lambda_{SOC}\lvert SOC_t-SOC^{ref}\rvert
-\lambda_H\lvert H_t-H^{ref}\rvert.
\label{eq:reward_function}
\end{equation}
Table~\ref{tab:reward_coefficients} identifies the coefficients inherited from the preceding PPO formulation \cite{atef2026deepreinforcementlearningbasedenergy} and their dimensional interpretation. The SOC and hydrogen-inventory terms are reward-shaping penalties rather than direct cash costs; consequently, training results are reported in cumulative reward units rather than Australian dollars. Forecasts affect the policy through the state rather than by altering the reward.

\begin{table}[H]
\centering
\caption{Reported operating-cost and reward-shaping coefficients.}
\label{tab:reward_coefficients}
\small
\begin{tabular*}{\textwidth}{@{\extracolsep{\fill}}llll@{}}
\toprule
Coefficient & Value & Required units & Interpretation \\
\midrule
$c_B$ & 0.01 & A\$/\si{\kilo\watt\hour} & Battery-throughput charge \\
$c_{DG}$ & 0.20 & A\$/\si{\kilo\watt\hour} & Backup-generation charge \\
$c_H^{prod}$ & 0.40 & A\$/\si{\kilogram} & Hydrogen-production charge \\
$c_H^{cons}$ & 2.00 & A\$/\si{\kilogram} & Hydrogen-consumption charge \\
$\lambda_{unmet}$ & 5.00 & A\$/\si{\kilo\watt\hour} & Unserved-energy penalty \\
$\lambda_{SOC}$ & 0.02 & reward units/p.u. & Battery-SOC shaping penalty \\
$\lambda_H$ & 0.01 & reward units/p.u. & Hydrogen-level shaping penalty \\
\bottomrule
\end{tabular*}
\end{table}

\subsection{Proximal Policy Optimization Controller}

PPO is an on-policy actor--critic algorithm that limits excessively large policy updates. With probability ratio $\rho_t=\pi_{\theta}(a_t|s_t)/\pi_{\theta_{old}}(a_t|s_t)$, the clipped surrogate objective is
\begin{equation}
L^{clip}(\theta)=\mathbb{E}_t\left[\min\left(\rho_t\widehat{A}_t,
\mathrm{clip}(\rho_t,1-\epsilon,1+\epsilon)\widehat{A}_t\right)\right],
\label{eq:ppo_objective}
\end{equation}
where $\widehat{A}_t$ is the generalized advantage estimate and $\epsilon=0.25$. The Gaussian actor and value critic each use two 256-neuron ReLU layers, consistent with the base study \cite{atef2026deepreinforcementlearningbasedenergy}.

\subsection{Training Environment}

The environment simulates chronological hourly operation using the component models and constraints in Section~\ref{sec:SystemDescription}. The inherited PPO settings are discount factor 0.995, generalized-advantage factor 0.95, experience horizon 4,096, 10 update epochs, mini-batch size 256, entropy weight 0.001, and random seed 42. The comparison uses 100 training episodes for forecast-enriched and non-predictive PPO\@. A perfect-forecast case is evaluated as an upper information benchmark, while no-battery operation provides a physical baseline.

\section{Case Study and Simulation Setup}
\label{sec:CaseStudy}

This section presents the Australian community microgrid case study and simulation environment used to evaluate the proposed forecasting and predictive energy management framework. Details regarding datasets, forecasting model configurations, DRL hyperparameters, and benchmark methods are provided.

\subsection{Australian Community Microgrid}

The case study represents a 1000-household Australian residential community microgrid based on the PV--wind--battery--hydrogen configuration developed in the earlier planning and EMS studies \cite{atef2026integrated,atef2025techno}. The associated AUPEC study established the real-time residential microgrid EMS and Australian operating context that precede the forecast-enriched extension presented here \cite{atef2025real}. The aggregated load is therefore calculated for $N_h=1000$ households using Eq.~\eqref{eq:community_load} or, when a base profile is used, the scaling relation in Eq.~\eqref{eq:scaled_load}. The infrastructure includes PV arrays, wind generation, BESS capacity, an electrolyzer, hydrogen storage, a fuel cell, and a grid interconnection. Renewable and storage capacities are treated as community-level assets sized for the 1000-household case, while grid import/export limits define the boundary between local operation and the upstream distribution network.

\subsection{Computational Environment} 

The proposed forecasting and predictive energy management framework was implemented in MATLAB using custom-developed scripts. The hydrogen-enabled community microgrid, including the PV system, wind generation, BESS, hydrogen subsystem, diesel backup generator, and grid interaction, was modeled within the MATLAB environment. The multi-horizon load forecasting outputs were incorporated as external inputs to the PPO-based energy management system. All simulations were conducted using hourly chronological data over one year (8,760 hours), and the operational, economic, and environmental performance metrics were evaluated within the same computational environment. The implementation was developed specifically for this study and does not rely on commercial microgrid simulation software.

\subsection{Input Datasets}
The dataset used in this study is derived from the Australian community microgrid case developed in our previous studies \cite{atef2026deepreinforcementlearningbasedenergy}. It comprises one year (8,760 hourly observations) of chronological data, including aggregated residential community load, available photovoltaic (PV) generation, available wind generation, grid purchase price, and grid export price. These data serve as inputs to both the forecasting model and the PPO-based energy management system. Table 3 summarizes the statistical characteristics of the input dataset.
%
Exogenous variables are normalized for neural-network inputs, while raw values are retained for power balancing, cost accounting, and performance evaluation. The forecast-result set reports a peak evaluation load of \SI{1684.51}{\kilo\watt}; this is lower than the \SI{2134.13}{\kilo\watt} maximum of the complete annual input record, indicating that the forecast evaluation used a subset or processed series rather than the raw annual maximum.

\begin{table}[H]
\centering
\caption{Summary of the 1000-household hourly input dataset.}
\label{tab:input_data}
\small
\begin{tabular*}{\textwidth}{@{\extracolsep{\fill}}lrrr@{}}
\toprule
Variable & Mean & Maximum & Annual total \\
\midrule
Community load & \SI{468.75}{\kilo\watt} & \SI{2134.13}{\kilo\watt} & \SI{4106249.99}{\kilo\watt\hour} \\
PV availability & \SI{1620.91}{\kilo\watt} & \SI{7973.49}{\kilo\watt} & \SI{14199183.53}{\kilo\watt\hour} \\
Wind availability & \SI{18.27}{\kilo\watt} & \SI{60.00}{\kilo\watt} & \SI{160045.69}{\kilo\watt\hour} \\
Grid purchase price & A\$0.3060/\si{\kilo\watt\hour} & A\$0.4571/\si{\kilo\watt\hour} & -- \\
Grid export price & A\$0.0601/\si{\kilo\watt\hour} & A\$0.0601/\si{\kilo\watt\hour} & -- \\
\bottomrule
\end{tabular*}
\end{table}

\subsection{Forecasting Evaluation Configuration}

The forecasting evaluation targets aggregated community load at hourly resolution over six forecast horizons. Table~\ref{tab:forecast_evidence} summarizes the evaluation settings used to report the available forecast results.

\begin{table}[H]
\centering
\caption{Forecasting evaluation configuration.}
\label{tab:forecast_evidence}
\small
\begin{tabular*}{\textwidth}{@{\extracolsep{\fill}}ll@{}}
\toprule
Item & Reported setting \\
\midrule
Forecast target & Aggregated community load \\
Time resolution & Hourly \\
Forecast horizons & 1, 6, 12, 24, 48, and 72 h \\
Uncertainty output & 95\% uncertainty interval for 24-h forecast \\
Accuracy metrics & RMSE, MAE, MAPE, nRMSE, correlation, $R^2$ \\
\bottomrule
\end{tabular*}
\end{table}

\subsection{Forecasting Training Settings}

The forecast series are generated independently of PPO and remain fixed during controller training, allowing forecast error to be evaluated separately from control performance.

\subsection{PPO Hyperparameters}

The PPO architecture and update parameters are inherited from the base DRL manuscript, while the forecast comparison uses 100 episodes. Table~\ref{tab:ppo_hyperparameters} lists the inherited settings. The convergence results also include a normalized learning-progress indicator of 0.82 for forecast-enriched PPO and 0.72 for non-predictive PPO; these values are not used as optimizer learning rates.

\begin{table}[H]
\centering
\caption{PPO settings inherited from the base hydrogen-microgrid EMS.}
\label{tab:ppo_hyperparameters}
\small
\setlength{\tabcolsep}{3pt}
\begin{tabular*}{\textwidth}{@{\extracolsep{\fill}}llll@{}}
\toprule
Parameter & Value & Parameter & Value \\
\midrule
Forecast comparison episodes & 100 & Discount factor & 0.995 \\
GAE factor & 0.95 & Experience horizon & 4,096 \\
Update epochs & 10 & Mini-batch size & 256 \\
PPO clip & 0.25 & Entropy weight & 0.001 \\
Actor network & $2\times256$ ReLU & Critic network & $2\times256$ ReLU \\
Battery scale & \SI{530}{\kilo\watt} & Diesel scale & \SI{180}{\kilo\watt} \\
\bottomrule
\end{tabular*}
\end{table}

The principal physical parameters are a \SI{5300}{\kilo\watt\hour} battery with 0.95 charge/discharge efficiency, a \SI{1370}{\kilogram} hydrogen tank, an electrolyzer limited to \SI{3.90}{\kilogram\per\hour} with 0.65 efficiency, and a fuel-cell withdrawal limit of \SI{7.50}{\kilogram\per\hour} with 0.60 efficiency \cite{atef2026deepreinforcementlearningbasedenergy}.

\subsection{Benchmark Methods}

Forecasting is compared with linear regression, ridge regression, lasso regression, an autoregressive model, and a time-of-day average. For energy management, four information/configuration cases are used: no battery, non-predictive PPO, forecast-enriched PPO, and perfect-forecast PPO\@. The non-predictive PPO case preserves the controller and physical model while removing future load information, making it the primary benchmark for the operational value of forecasting. The perfect-forecast case defines an upper information bound rather than a deployable controller. The no-battery case quantifies the combined value of storage and control.

\section{Results}
\label{sec:Results}

This section presents and analyzes the forecasting and predictive energy management performance of the proposed framework. The forecasting accuracy and resulting impacts on microgrid operation are evaluated from economic, technical, environmental, and resilience perspectives.

\subsection{Forecasting Performance}

Table~\ref{tab:forecast_metrics_results} summarizes the multi-horizon community-load forecasts. The 1-h forecast provides the lowest RMSE (\SI{239.32}{\kilo\watt}) and nRMSE (0.114), with correlation 0.725 and $R^2=0.201$. Accuracy deteriorates sharply at 6 and 12 h: $R^2$ becomes $-0.757$ and $-1.914$, respectively, and the 12-h correlation is negative. The model therefore performs worse than a mean-value predictor at these intermediate horizons. At 24--72 h, RMSE returns to approximately \SIrange{246}{250}{\kilo\watt} and correlation stabilizes near 0.686. No causal interpretation is assigned to this non-monotonic horizon pattern without timestamp-alignment and sample-count information. MAPE is interpreted cautiously because low demand values can make its denominator unstable; the value of 126.37\% at 12 h should not be treated as a scale-independent summary of accuracy.

\begin{table}[H]
\centering
\caption{Multi-horizon community-load forecasting performance.}
\label{tab:forecast_metrics_results}
\footnotesize
\setlength{\tabcolsep}{3.5pt}
\begin{tabular}{@{}rrrrrrr@{}}
\toprule
Horizon (\si{\hour}) & RMSE (\si{\kilo\watt}) & MAE (\si{\kilo\watt}) & MAPE (\%) & $R^2$ & Correlation & nRMSE \\
\midrule
1  & 239.32 & 190.90 & 58.87  & 0.201  & 0.725  & 0.114 \\
6  & 354.10 & 287.13 & 84.25  & -0.757 & 0.208  & 0.169 \\
12 & 456.48 & 385.24 & 126.37 & -1.914 & -0.423 & 0.218 \\
24 & 249.79 & 197.36 & 62.52  & 0.126  & 0.683  & 0.119 \\
48 & 247.62 & 195.54 & 61.92  & 0.136  & 0.687  & 0.118 \\
72 & 246.46 & 193.85 & 61.22  & 0.143  & 0.686  & 0.118 \\
\bottomrule
\end{tabular}
\end{table}

\begin{figure}[H]
\centering
\includegraphics[width=0.98\textwidth]{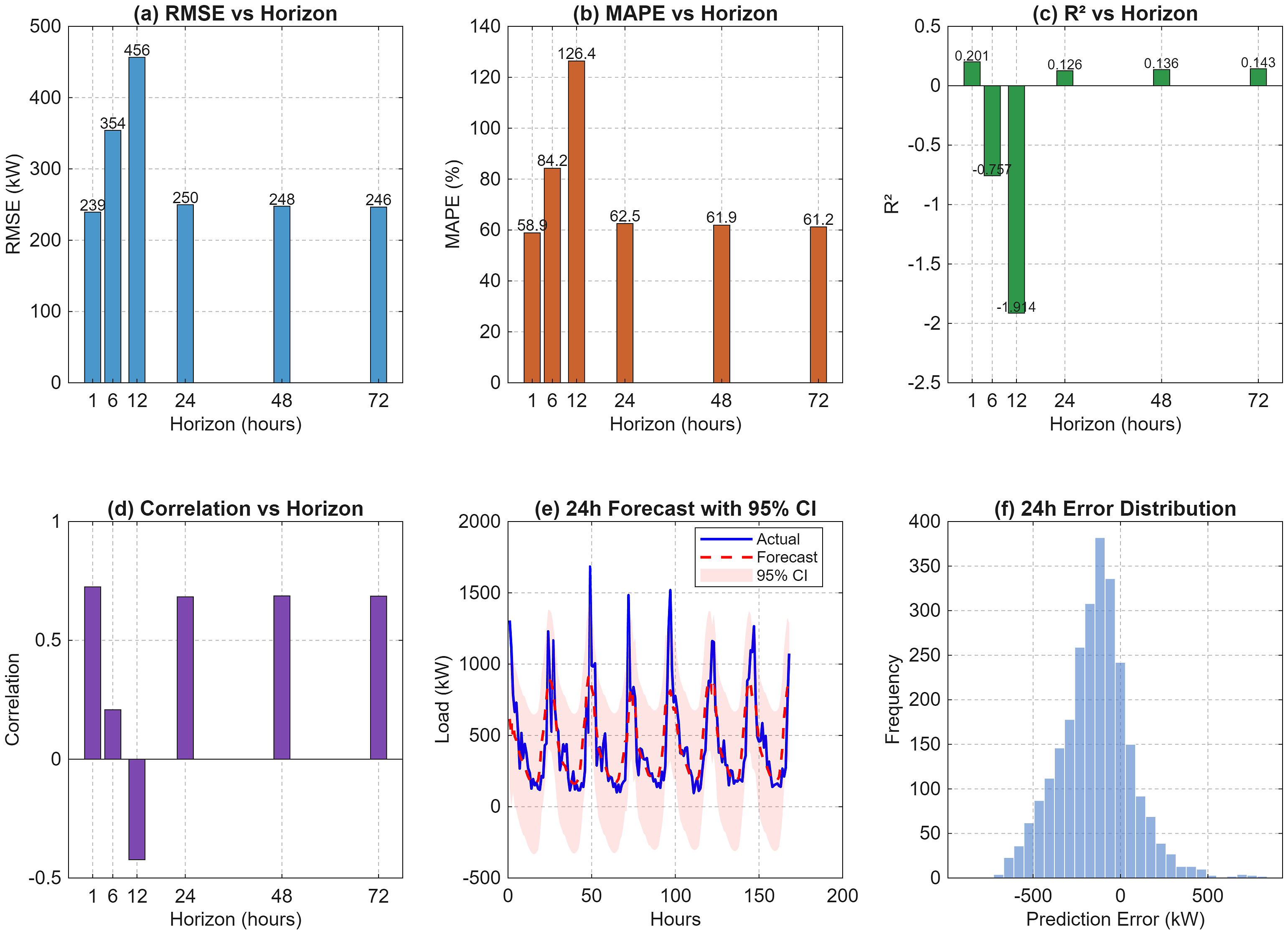}
\caption{Forecasting results across prediction horizons, including the 24-h uncertainty interval and error distribution.}
\label{fig:forecast_results_original}
\end{figure}

Figure~\ref{fig:forecast_results_original} shows the horizon dependence, the 24-h prediction with its reported 95\% uncertainty interval, and the error distribution. The forecast follows the recurring load cycle but underestimates several sharp peaks. The error histogram is concentrated below zero, indicating a tendency toward underprediction.

\subsection{Forecasting Model Comparison}

A forecasting-model ranking is outside the scope of this operational-value study because the available baseline scores were not generated under the same horizon-specific evaluation conditions as Table~\ref{tab:forecast_metrics_results}. Accordingly, no superiority claim is made. A future forecasting comparison must use identical chronological samples, forecast origins, horizons, input information, scaling, and error definitions.

\subsection{DRL Training Performance}

The single-seed training trajectories in Table~\ref{tab:training_comparison} show higher observed reward for forecast-enriched PPO\@. The forecast-enriched controller begins at 24,500 reward units and reaches 27,571, compared with 23,800 and 25,455 for non-predictive PPO\@. The final-episode difference is 8.3\%. Visual stabilization occurs near Episode 30 for forecast-enriched PPO and Episode 35 for non-predictive PPO; these episode numbers are descriptive markers rather than statistically validated convergence times.

\begin{table}[H]
\centering
\caption{PPO training performance for forecast-enriched and non-predictive control.}
\label{tab:training_comparison}
\small
\setlength{\tabcolsep}{3pt}
\begin{tabular*}{\textwidth}{@{\extracolsep{\fill}}>{\raggedright\arraybackslash}p{0.31\textwidth}rrr@{}}
\toprule
Metric & \shortstack{Forecast-enriched\\PPO} & \shortstack{Non-predictive\\PPO} & \shortstack{Observed\\difference} \\
\midrule
Initial reward (reward units) & 24,500 & 23,800 & 2.9\% \\
Final reward (reward units) & 27,571 & 25,455 & 8.3\% \\
Training reward increase (reward units) & 3,071 & 1,655 & 85.6\% \\
Visual stabilization episode & 30 & 35 & 14.3\% earlier \\
Normalized learning indicator & 0.82 & 0.72 & 13.9\% \\
\bottomrule
\end{tabular*}
\end{table}

\begin{figure}[H]
\centering
\includegraphics[width=0.98\textwidth]{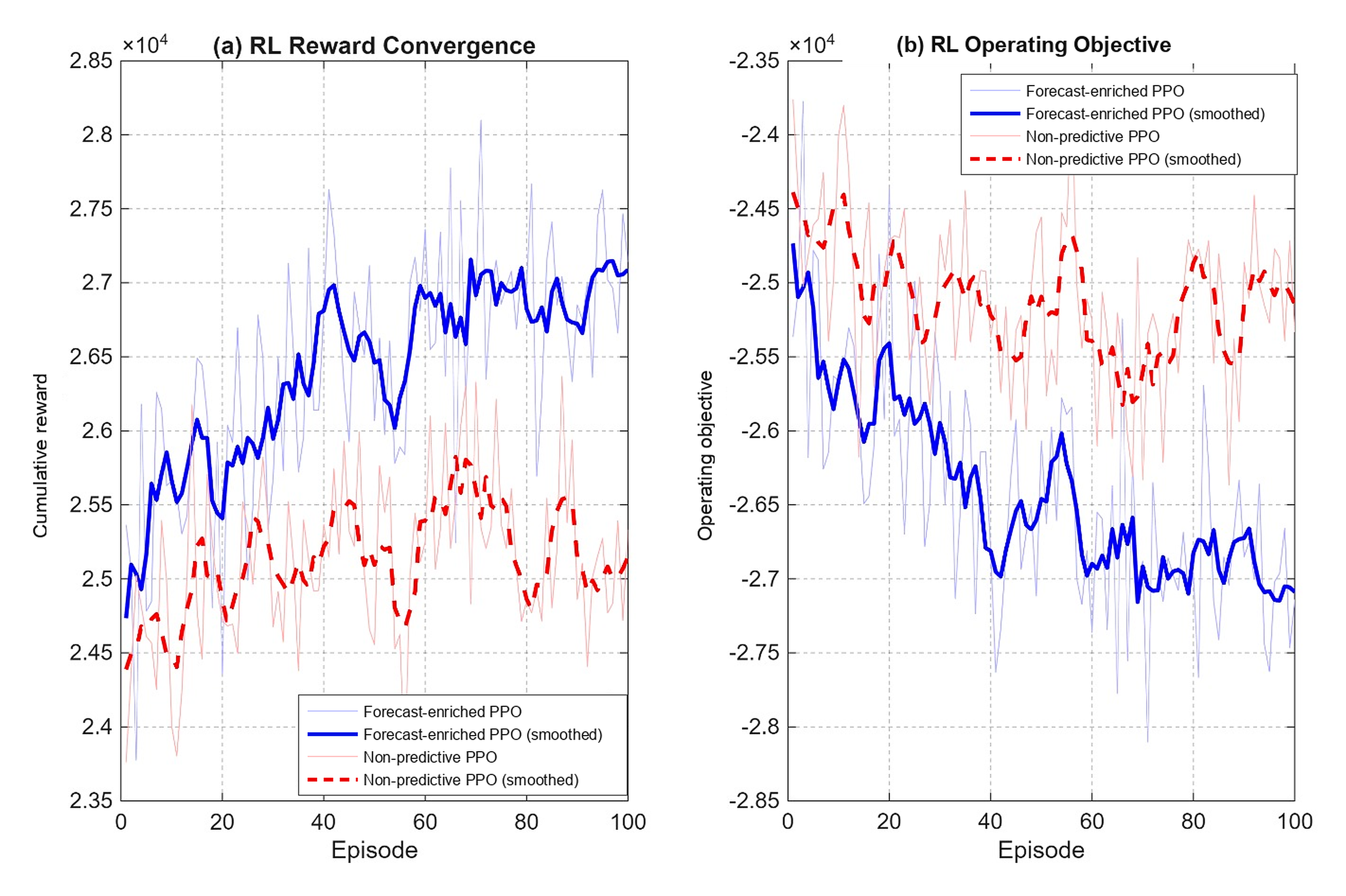}
\caption{Cumulative PPO reward and operating-objective traces for forecast-enriched and non-predictive control in the reported training seed.}
\label{fig:ppo_convergence_original}
\end{figure}

Figure~\ref{fig:ppo_convergence_original} shows that the smoothed forecast-enriched reward remains above the non-predictive reward after the initial training period in the reported seed. The episode-to-episode traces remain noisy, so this figure provides descriptive training evidence rather than a multi-seed statistical comparison.

\subsection{Impact of Forecasting on Energy Management}

The forecast-enriched policy produces A\$2,765.83 in annual savings, compared with A\$2,439.86 for non-predictive PPO (Table~\ref{tab:ems_savings}). The incremental value of forecast information is A\$325.97, equivalent to 13.4\% relative to the non-predictive savings. The perfect-forecast case reaches A\$3,054.94. Relative to non-predictive PPO, the implemented forecast therefore captures approximately 53.0\% of the additional economic benefit available under perfect load information. The incremental benefit corresponds to A\$0.326 per household per year and A\$0.079 per MWh of annual community demand. These values quantify aggregate operating benefits only and do not include the cost of deploying or maintaining the forecasting infrastructure.

\begin{table}[H]
\centering
\caption{Operational value of forecast information.}
\label{tab:ems_savings}
\small
\begin{tabular*}{\textwidth}{@{\extracolsep{\fill}}lrr@{}}
\toprule
Scenario & Savings (A\$) & Savings (\%) \\
\midrule
Forecast-enriched PPO & 2,765.83 & 11.2 \\
Non-predictive PPO & 2,439.86 & 9.9 \\
Perfect-forecast PPO & 3,054.94 & 12.4 \\
\bottomrule
\end{tabular*}
\end{table}

\subsection{Operational Performance}

The main operational effects are summarized in Table~\ref{tab:operational_results}. Relative to no battery, forecast-enriched PPO reduces peak net load from 681.79 to \SI{676.26}{\kilo\watt} and grid imports from 59,594.33 to \SI{58147.49}{\kilo\watt\hour}. Relative to non-predictive PPO, the incremental import reduction is \SI{170.15}{\kilo\watt\hour} (0.29\%), and peak net load falls by \SI{0.65}{\kilo\watt}.

\begin{table}[H]
\centering
\caption{Operational performance metrics.}
\label{tab:operational_results}
\footnotesize
\setlength{\tabcolsep}{3pt}
\begin{tabular*}{\textwidth}{@{\extracolsep{\fill}}>{\raggedright\arraybackslash}p{0.34\textwidth}rrr@{}}
\toprule
Metric & No battery & \shortstack{Forecast-enriched\\PPO} & \shortstack{Non-predictive\\PPO} \\
\midrule
Peak load (\si{\kilo\watt}) & 1,684.51 & 1,684.51 & 1,684.51 \\
Peak net load (\si{\kilo\watt}) & 681.79 & 676.26 & 676.91 \\
Maximum ramp rate (\si{\kilo\watt\per\hour}) & 1,165.09 & 1,165.09 & 1,165.09 \\
Curtailment (\si{\kilo\watt\hour}) & 762,274.21 & 763,593.19 & 763,437.38 \\
Grid import (\si{\kilo\watt\hour}) & 59,594.33 & 58,147.49 & 58,317.64 \\
Grid export (\si{\kilo\watt\hour}) & 760,214.31 & 764,243.18 & 764,387.58 \\
\bottomrule
\end{tabular*}
\end{table}

\subsection{Environmental Performance}

Direct grid-import emissions are calculated as
\begin{equation}
E_{grid}=e_{grid}\sum_t P_{grid,t}^{buy}\Delta t,
\label{eq:grid_emissions}
\end{equation}
using the constant simulation factor $e_{grid}=\SI{0.5}{\kilogram\per\kilo\watt\hour}$ CO$_2$. Forecast-enriched PPO reduces grid imports by 2.4\% relative to the no-battery case, lowering direct grid-import emissions from 29,797.17 to \SI{29073.74}{\kilogram} CO$_2$. Non-predictive PPO produces \SI{29158.82}{\kilogram} CO$_2$; forecast information therefore reduces direct import-related emissions by \SI{85.08}{\kilogram} CO$_2$.

\begin{table}[H]
\centering
\caption{Direct grid-import emissions under the constant simulation emissions factor.}
\label{tab:environmental_results}
\footnotesize
\setlength{\tabcolsep}{3pt}
\begin{tabular*}{\textwidth}{@{\extracolsep{\fill}}>{\raggedright\arraybackslash}p{0.36\textwidth}rrr@{}}
\toprule
Metric & No battery & \shortstack{Forecast-enriched\\PPO} & \shortstack{Non-predictive\\PPO} \\
\midrule
Grid import (\si{\kilo\watt\hour}) & 59,594.33 & 58,147.49 & 58,317.64 \\
Direct grid-import emissions (\si{\kilogram} CO$_2$) & 29,797.17 & 29,073.74 & 29,158.82 \\
\bottomrule
\end{tabular*}
\end{table}

\subsection{Resilience Scope}

No forecast-specific resilience benefit is claimed from the available results. A defensible resilience comparison requires physical reliability metrics such as energy not served, loss-of-load hours, minimum battery SOC, minimum hydrogen inventory, and supported outage duration under matched initial conditions.

\subsection{Comparative Analysis}

The consolidated comparison in Table~\ref{tab:overall_comparison} summarizes the aggregate economic and grid-exchange outcomes retained in the revised analysis.

\begin{table}[H]
\centering
\caption{Consolidated comparison of the principal cases.}
\label{tab:overall_comparison}
\footnotesize
\setlength{\tabcolsep}{3pt}
\begin{tabular*}{\textwidth}{@{\extracolsep{\fill}}>{\raggedright\arraybackslash}p{0.36\textwidth}rrr@{}}
\toprule
Metric & No battery & \shortstack{Forecast-enriched\\PPO} & \shortstack{Non-predictive\\PPO} \\
\midrule
Economic savings (A\$) & 0.00 & 2,765.83 & 2,439.86 \\
Grid import (\si{\kilo\watt\hour}) & 59,594.33 & 58,147.49 & 58,317.64 \\
Peak net load (\si{\kilo\watt}) & 681.79 & 676.26 & 676.91 \\
Direct grid-import emissions (\si{\kilogram} CO$_2$) & 29,797.17 & 29,073.74 & 29,158.82 \\
\bottomrule
\end{tabular*}
\end{table}

\section{Discussion}
\label{sec:Discussion}

The findings are interpreted as proof-of-concept evidence for the operational value of forecast information, rather than as validation of a forecasting architecture or a statistically robust PPO advantage. This distinction is necessary because the forecast series are treated as external inputs, the forecasting benchmarks are not directly comparable, and the PPO results represent one training seed.

\subsection{Forecast Accuracy and Horizon Dependence}

The forecasting results are strongly horizon dependent. At 1 h, the correlation of 0.725 indicates that the forecast tracks much of the short-term temporal movement, but $R^2=0.201$ shows that most load variance remains unexplained. The corresponding RMSE of \SI{239.32}{\kilo\watt} is substantial relative to the mean community demand of \SI{468.75}{\kilo\watt}. Therefore, even the best reported horizon should be regarded as informative rather than highly accurate.

The 6- and 12-h results are materially weaker. Negative $R^2$ values mean that these forecasts perform worse than predicting the evaluation-set mean under the coefficient-of-determination criterion. The 12-h correlation of $-0.423$ is particularly concerning because it indicates inverse temporal association, not merely amplitude error. A controller receiving these inputs could make poorly timed reserve decisions unless it learns to suppress their influence. Conversely, PPO may learn to ignore unreliable state variables, which could explain why aggregate operation improves despite poor intermediate-horizon forecasts. This explanation remains hypothetical without a horizon-ablation study or policy-sensitivity analysis.

The recovery at 24--72 h, where correlation remains near 0.686 and RMSE returns to approximately 246--250 kW, suggests that the forecasts retain a broad periodic load pattern. However, the near-identical metrics at 24, 48, and 72 h are unusual enough that they should not be interpreted as evidence that multi-day prediction is as reliable as day-ahead prediction. Timestamp alignment, sample counts, target shifting, and inverse normalization must be verified before attributing this pattern to genuine daily periodicity. The high MAPE values are also interpreted cautiously because observations near zero can dominate percentage errors. Finally, the tendency to underpredict peaks is operationally important: underestimated demand can leave insufficient stored energy before high-load periods, even when average forecasting metrics appear acceptable.

\subsection{PPO Learning Performance}

Forecast-enriched PPO reaches a final observed reward 8.3\% above non-predictive PPO in the reported trajectory. The separation between the smoothed curves after the early episodes is consistent with the possibility that future-load information helps the policy identify more favorable storage and grid actions. The larger training reward increase of 3,071 reward units, compared with 1,655 units for non-predictive PPO, points in the same direction.

These differences should not be treated as a robust convergence result. The forecast-enriched trajectory already begins 2.9\% above the non-predictive trajectory, so not all of the final separation necessarily arises from faster learning. Episode-to-episode noise is large, the comparison uses one seed, and the final episode may not represent the expected policy performance. The visually identified Episodes 30 and 35 are therefore stabilization markers rather than formal convergence times. A defensible RL comparison requires paired random seeds, common initial storage states, mean reward over a final evaluation window, and deterministic out-of-sample policy testing. In addition, the reported reward includes shaping penalties; its magnitude is useful for comparing training behavior but is not itself an Australian-dollar economic outcome.

\subsection{Economic Value of Forecast Information}

The forecast-enriched policy increases annual savings by A\$325.97 relative to non-predictive PPO\@. The 13.4\% relative improvement appears sizeable because it is calculated against the smaller non-predictive savings of A\$2,439.86. In absolute terms, however, the gain is only A\$0.326 per household per year and A\$0.079 per MWh of annual community demand. The percentage and absolute values therefore convey different practical messages: forecasting improves the reported control benefit, but the community-scale financial impact is modest.

The perfect-forecast benchmark provides a useful value-of-information bound. The additional benefit available between non-predictive and perfect-forecast PPO is A\$615.08, of which the implemented forecasts capture A\$325.97, or approximately 53.0\%. This result indicates both useful forecast value and substantial remaining opportunity. It does not identify whether the remaining gap is caused by forecast error, PPO training variability, reward design, or restrictions in the physical system.

The economic interpretation is also limited by the absence of component-level hourly cash flows. Aggregate savings cannot reveal whether the improvement originates from reduced import expenditure, altered export timing, battery throughput, or hydrogen conversion. The direct financial benefit may not justify a dedicated forecasting platform unless forecasting already exists within the EMS or supports additional services such as demand-charge reduction, flexibility-market participation, or constrained-grid management.

\subsection{Operational Effects}

The operational differences between forecast-enriched and non-predictive PPO are small. Grid imports decrease by only \SI{170.15}{\kilo\watt\hour}, or 0.29\%, and peak net load falls by \SI{0.65}{\kilo\watt}. Peak gross load and maximum ramp rate are unchanged, which is expected because load forecasting changes dispatch decisions rather than the underlying demand profile.

Other indicators do not show a uniformly favorable physical effect. Curtailment is \SI{155.81}{\kilo\watt\hour} higher under forecast-enriched control, while grid export is \SI{144.40}{\kilo\watt\hour} lower. These changes are very small relative to annual renewable availability, but they show that the economic gain is not produced by broad reductions in curtailed energy or by substantially greater export. The most plausible interpretation is that forecasts alter the timing of transactions and storage actions rather than the annual energy quantities. Component-level battery and hydrogen trajectories are required to test this mechanism directly.

\subsection{Environmental Implications}

The direct emissions result follows mechanically from the change in grid imports because a constant factor of \SI{0.5}{\kilogram\per\kilo\watt\hour} CO$_2$ is applied. The forecast-specific reduction of \SI{85.08}{\kilogram} CO$_2$ is approximately 0.29\% of the non-predictive grid-import emissions, matching the percentage reduction in imported energy. Thus, the environmental benefit is positive but small and does not constitute an independent effect beyond the grid-import reduction.

The constant emissions factor also prevents the analysis from valuing the timing of imports. Forecast-assisted scheduling could have a larger or smaller carbon effect if marginal grid emissions vary hourly, but that cannot be established here. Diesel emissions, battery lifecycle emissions, and hydrogen-production emissions are not included in the retained direct-emissions boundary. The environmental conclusion is therefore restricted to operational emissions from imported electricity.

\subsection{Hydrogen and Resilience Evidence}

The mathematical formulation represents a hydrogen-enabled microgrid, but the reported aggregate results do not demonstrate how forecasts affect electrolyzer loading, hydrogen production, tank inventory, or fuel-cell dispatch. Consequently, the observed economic difference cannot be attributed specifically to improved hydrogen coordination. The same limitation applies to the battery because annual aggregate metrics do not identify when or why storage actions change.

No forecast-specific resilience conclusion is supported by the retained evidence. Resilience should be assessed through physical metrics such as energy not served, loss-of-load hours, minimum SOC, minimum hydrogen inventory, critical-load service, and supported outage duration under matched starting conditions. Removing the inconsistent aggregate stress-cost table avoids overstating resilience, but it also means that resilience remains an open experimental question rather than a demonstrated contribution.

\subsection{Practical Deployment and Generalizability}

Once trained, PPO inference requires a neural-network forward pass and is compatible with hourly supervisory control. Forecast generation and PPO inference can be executed sequentially, while the feasibility mapping clips actions, enforces operating modes, and reconciles the power balance. Safe DRL and hybrid MPC--DRL studies provide suitable directions for strengthening this deployment layer \cite{wang2025safe,liu2025novel}.

Generalization beyond the Rockhampton case is not established. The study uses one 1000-household community, one year of data, one tariff structure, and one reported PPO seed. Transfer to another community would require local load and weather data, retraining, rescaled DER and storage capacities, and validation under the local grid and tariff constraints. Before deployment or strong journal claims, the analysis should also confirm chronological separation between forecasting and PPO evaluation, define the exact forecast vector supplied to the controller, and quantify performance across multiple seeds and forecast-error conditions.

\section{Conclusion}
\label{sec:Conclusion}

This paper evaluated the operational value of multi-horizon community-load forecasts in a PPO-based EMS for a 1000-household hydrogen-enabled community microgrid. Forecasting performance is mixed: the 1-h horizon achieves an RMSE of \SI{239.32}{\kilo\watt} and $R^2=0.201$, while the 6-h and 12-h horizons perform poorly. In the reported single-seed experiment, forecast-enriched PPO reaches a final reward 8.3\% above non-predictive PPO and increases annual savings from A\$2,439.86 to A\$2,765.83. This A\$325.97 incremental benefit equals A\$0.326 per household per year and captures approximately 53.0\% of the additional economic benefit available under perfect load information. The evidence supports improved economic dispatch under the tested simulation, but it does not establish forecasting superiority, multi-seed RL robustness, forecast-enhanced resilience, or improved hydrogen coordination.

\bibliographystyle{elsarticle-num}
\bibliography{references}

@article{atef2026energy,
  title={Energy Management in Microgrids: Commercial, Industrial, and Residential Perspectives},
  author={Atef, Mohamed and Alahakoon, Sanath and Wolfs, Peter and Mumtahina, Umme and Khatib, Tamer and Uddin, Moslem},
  journal={Energies},
  volume={19},
  number={2},
  pages={419},
  year={2026},
  publisher={MDPI}
}

@article{atef2026review,
  title={A review of microgrid energy management systems: methods, challenges, and future directions},
  author={Atef, Mohamed and Alahakoon, Sanath and Mumtahina, Umme and Wolfs, Peter and Khatib, Tamer and Uddin, Moslem},
  journal={International Journal of Ambient Energy},
  volume={47},
  number={1},
  pages={2595117},
  year={2026},
  publisher={Taylor \& Francis}
}

@inproceedings{atef2019utilization,
  title={Utilization of artificial neural networks to improve the accuracy of a hybrid power system model},
  author={Atef, Mohamed and Abdullah, MF and Khatib, Tamer and Romlie, MF},
  booktitle={2019 IEEE Student Conference on Research and Development (SCOReD)},
  pages={284--288},
  year={2019},
  organization={IEEE}
}

@article{atef2020optimization,
  title={Optimization of a Hybrid Solar {PV} and Gas Turbine Generator System Using the Loss of Load Probability Index},
  author={Atef, Mohamed and Khatib, Tamer and Abdullah, M. F. and Romlie, M. F.},
  journal={Clean Technologies},
  volume={2},
  number={3},
  pages={240--251},
  year={2020},
  publisher={MDPI},
  doi={10.3390/cleantechnol2030016}
}

@incollection{atef2024datadriven,
  title={Data-driven optimization framework for microgrid energy management},
  author={Atef, Mohamed and Uddin, Moslem and Rana, Md Masud and Sarkar, Md Rasel and Shafiullah, G. M.},
  booktitle={Intelligent Data-Driven Modelling and Optimization in Power and Energy Applications},
  pages={169--188},
  year={2024},
  publisher={CRC Press},
  doi={10.1201/9781003470274-8}
}

@inproceedings{atef2025techno,
  title={Techno-Economic Energy Planning of Residential Microgrids: An Australian Case Study},
  author={Atef, Mohamed and Alahakoon, Sanath and Wolfs, Peter and Khatib, Tamer and Mumtahina, Umme and Uddin, Moslem},
  booktitle={2025 IEEE PES 35th Australasian Universities Power Engineering Conference (AUPEC)},
  pages={1--5},
  year={2025},
  organization={IEEE}
}

@inproceedings{atef2025real,
  title={Real-Time Energy Management of Residential Microgrids: An Australian Case Study},
  author={Atef, Mohamed and Alahakoon, Sanath and Wolfs, Peter and Khatib, Tamer and Mumtahina, Umme and Uddin, Moslem},
  booktitle={2025 IEEE PES 35th Australasian Universities Power Engineering Conference (AUPEC)},
  pages={1--5},
  year={2025},
  organization={IEEE},
  doi={10.1109/AUPEC66173.2025.11219602}
}

@article{modu2023systematic,
  title={A systematic review of hybrid renewable energy systems with hydrogen storage: Sizing, optimization, and energy management strategy},
  author={Modu, Babangida and Abdullah, Md Pauzi and Bukar, Abba Lawan and Hamza, Mukhtar Fatihu},
  journal={International Journal of Hydrogen Energy},
  year={2023},
  publisher={Elsevier}
}

@article{elkhatib2024green,
  title={Green hydrogen energy source for a residential fuel cell micro-combined heat and power},
  author={Elkhatib, Rafik and Kaoutari, Taoufiq and Louahlia, Hasna},
  journal={Applied Thermal Engineering},
  volume={248},
  pages={123194},
  year={2024},
  publisher={Elsevier}
}

@article{das2024sustainable,
  title={Sustainable integration of green hydrogen in renewable energy systems for residential and EV applications},
  author={Das, Deboleena and Chakraborty, Ishan and Bohre, Aashish Kumar and Kumar, Prince and Agarwala, Raja},
  journal={International Journal of Energy Research},
  volume={2024},
  number={1},
  pages={8258624},
  year={2024},
  publisher={Wiley Online Library}
}

@article{enaloui2025techno,
  title={Techno-economic assessment of a solar-powered green hydrogen storage concept based on reversible solid oxide cells for residential micro-grid: A case study in Calgary},
  author={Enaloui, Reza and Sharifi, Shakiba and Faridpak, Behdad and Hammad, Ahmed and Al-Hussein, Mohamed and Musilek, Petr},
  journal={Energy},
  volume={319},
  pages={134981},
  year={2025},
  publisher={Elsevier}
}

@article{Witharama2024,
  author    = {Witharama, W. M. N. and Bandara, K. and Bandara, K. M. D. P. and Azeez, M. I. and Wanigasekara, C. and Logeeshan, V.},
  title     = {Advanced Genetic Algorithm for Optimal Microgrid Scheduling Considering Solar and Load Forecasting, Battery Degradation, and Demand Response Dynamics},
  journal   = {IEEE Access},
  year      = {2024},
  volume    = {12},
  pages     = {83269--83284},
  doi       = {10.1109/access.2024.3412914}
}

@article{Houben2023,
  author    = {Houben, N. and Cosic, A. and Stadler, M. and Mansoor, M. and Zellinger, M. and Auer, H. and Ajanovic, A. and Haas, R.},
  title     = {Optimal dispatch of a multi-energy system microgrid under uncertainty: A renewable energy community in Austria},
  journal   = {Applied Energy},
  year      = {2023},
  volume    = {337},
  pages     = {120913},
  doi       = {10.1016/j.apenergy.2023.120913}
}

@article{Majeed2023,
  author    = {Majeed, M. A. and Asghar, F. and Hussan, U. and Phichaisawat, S.},
  title     = {Optimal Energy Management System for Grid-Tied Microgrid: An Improved Adaptive Genetic Algorithm},
  journal   = {IEEE Access},
  year      = {2023},
  volume    = {11},
  pages     = {117351--117361},
  doi       = {10.1109/access.2023.3326505}
}

@article{Niknami2024,
  author    = {Niknami, A. and Tolou Askari, M. and Amir Ahmadi, M. and Babaei Nik, M. and Samiei Moghaddam, M.},
  title     = {Resilient day-ahead microgrid energy management with uncertain demand, EVs, storage, and renewables},
  journal   = {Cleaner Engineering and Technology},
  year      = {2024},
  volume    = {20},
  pages     = {100763},
  doi       = {10.1016/j.clet.2024.100763}
}

@article{Tang2024,
  author    = {Tang, W. and Cao, L. and Chen, Y. and Chen, B. and Yue, Y.},
  title     = {Solving Engineering Optimization Problems Based on Multi-Strategy Particle Swarm Optimization Hybrid Dandelion Optimization Algorithm},
  journal   = {Biomimetics (Basel, Switzerland)},
  year      = {2024},
  volume    = {9},
  number    = {5},
  pages     = {298},
  doi       = {10.3390/biomimetics9050298}
}

@article{Asif2024,
  author    = {Asif, M. and Razzaq, S. and Amin, A. and Mahdi, F. P. and Ahmed, U. and Jamil, U. and Mahmood, A.},
  title     = {Combined emission economic dispatch using quantum-inspired particle swarm optimization and its variants},
  journal   = {Energy Exploration \& Exploitation},
  year      = {2024},
  volume    = {42},
  number    = {5},
  pages     = {1602--1644},
  doi       = {10.1177/01445987241235419}
}

@article{Bhuvaneshwarri2025,
  author    = {Bhuvaneshwarri, B. and Kalaivanan, C. and Kanth, T. and Deepthi, P. and Maheswari, M. and Saravanan, V.},
  title     = {Hybrid Swarm Intelligence-Based Neural Framework for Optimizing Real-Time Computational Models in Engineering Systems},
  journal   = {International Journal of Computational and Experimental Science and Engineering},
  year      = {2025},
  volume    = {11},
  number    = {1},
  doi       = {10.22399/ijcesen.1001}
}

@article{cabrera2022review,
  title={A Review of the Optimization and Control Techniques in the Presence of Uncertainties for the Energy Management of Microgrids},
  author={Cabrera-Tobar, Ana and Massi Pavan, Alessandro and Petrone, Giovanni and Spagnuolo, Giovanni},
  journal={Energies},
  volume={15},
  number={23},
  pages={9114},
  year={2022},
  publisher={MDPI}
}

@article{sharma2022critical,
  title={A critical and comparative review of energy management strategies for microgrids},
  author={Sharma, Pavitra and Mathur, Hitesh Dutt and Mishra, Puneet and Bansal, Ramesh C},
  journal={Applied Energy},
  volume={327},
  pages={120028},
  year={2022},
  publisher={Elsevier}
}

@article{allwyn2023comprehensive,
  title={A comprehensive review on energy management strategy of microgrids},
  author={Allwyn, Rona George and Al-Hinai, Amer and Margaret, Vijaya},
  journal={Energy Reports},
  volume={9},
  pages={5565--5591},
  year={2023},
  publisher={Elsevier}
}

@article{uddin2023microgrids,
  title={Microgrids: A review, outstanding issues and future trends},
  author={Uddin, Moslem and Mo, Huadong and Dong, Daoyi and Elsawah, Sondoss and Zhu, Jianguo and Guerrero, Josep M},
  journal={Energy Strategy Reviews},
  volume={49},
  pages={101127},
  year={2023},
  publisher={Elsevier}
}

@article{chartier2022microgrid,
	title={Microgrid emergence, integration, and influence on the future energy generation equilibrium—A Review},
	author={Chartier, Sabrina Lee and Venkiteswaran, Vinod Kumar and Rangarajan, Shriram S and Collins, Edward Randolph and Senjyu, Tomonobu},
	journal={Electronics},
	volume={11},
	number={5},
	pages={791},
	year={2022},
	publisher={MDPI}
}

@article{trivedi2022community,
	title={Community-based microgrids: Literature review and pathways to decarbonise the local electricity network},
	author={Trivedi, Rohit and Patra, Sandipan and Sidqi, Yousra and Bowler, Benjamin and Zimmermann, Fiona and Deconinck, Geert and Papaemmanouil, Antonios and Khadem, Shafi},
	journal={Energies},
	volume={15},
	number={3},
	pages={918},
	year={2022},
	publisher={MDPI}
}

@article{yang2022modelling,
  title={Modelling and optimal energy management for battery energy storage systems in renewable energy systems: A review},
  author={Yang, Yuqing and Bremner, Stephen and Menictas, Chris and Kay, Merlinde},
  journal={Renewable and Sustainable Energy Reviews},
  volume={167},
  pages={112671},
  year={2022},
  publisher={Elsevier}
}

@article{zia2022energy,
  title={Energy management system for a hybrid PV-Wind-Tidal-Battery-based islanded DC microgrid: Modeling and experimental validation},
  author={Zia, Muhammad Fahad and Nasir, Mashood and Elbouchikhi, Elhoussin and Benbouzid, Mohamed and Vasquez, Juan C and Guerrero, Josep M},
  journal={Renewable and Sustainable Energy Reviews},
  volume={159},
  pages={112093},
  year={2022},
  publisher={Elsevier}
}

@article{bilal2024review,
  title={Review of computational intelligence approaches for microgrid energy management},
  author={Bilal, Mohd and Algethami, Abdullah A and Hameed, Salman and others},
  journal={IEEE Access},
  year={2024},
  publisher={IEEE}
}

@article{mannan2024recent,
  title={Recent development of grid-connected microgrid scheduling controllers for sustainable energy: A bibliometric analysis and future directions},
  author={Mannan, M and Mansor, M and Reza, MS and Roslan, MF and Ker, Pin Jern and Hannan, MA},
  journal={IEEE Access},
  year={2024},
  publisher={IEEE}
}

@article{sadeghian2024energy,
  title={Energy management of hybrid fuel cell and renewable energy based systems-A review},
  author={Sadeghian, Omid and Shotorbani, Amin Mohammadpour and Ghassemzadeh, Saeid and Mohammadi-Ivatloo, Behnam},
  journal={International Journal of Hydrogen Energy},
  year={2024},
  publisher={Elsevier}
}

@article{atef2026integrated,
  title={An Integrated Techno-Economic Framework for Optimal Microgrid Design: An Australian Case Study},
  author={Atef, Mohamed and Alahakoon, Sanath and Mumtahina, Umme and Wolfs, Peter and Khatib, Tamer and Uddin, Moslem},
  journal={arXiv preprint arXiv:2606.03783},
  year={2026}
}

@misc{atef2026deepreinforcementlearningbasedenergy,
  title={Deep Reinforcement Learning-Based Energy Management for Hydrogen-Enabled Community Microgrids Under Uncertainty},
  author={Mohamed Atef and Sanath Alahakoon and Umme Mumtahina and Peter Wolfs and Tamer Khatib and Moslem Uddin},
  year={2026},
  eprint={2607.17248},
  archivePrefix={arXiv},
  primaryClass={eess.SY},
  url={https://arxiv.org/abs/2607.17248}
}

@article{quispe2025energy,
  title={Energy management systems in higher education institutions’ buildings},
  author={Quispe, Enrique C and Viveros Mira, Miguel and Chamorro D{\'\i}az, Mauricio and Castrill{\'o}n Mendoza, Rosaura and Vidal Medina, Juan R},
  journal={Energies},
  volume={18},
  number={7},
  pages={1810},
  year={2025},
  publisher={MDPI}
}

@article{ramkumar2025optimal,
  title={Optimal energy management for multi-energy microgrids using hybrid solutions to address renewable energy source uncertainty},
  author={Ramkumar, M Siva and Subramani, Jaganathan and Sivaramkrishnan, M and Munimathan, Arunkumar and Michael, Goh Kah Ong and Alam, Mohammad Mukhtar},
  journal={Scientific Reports},
  volume={15},
  number={1},
  pages={7755},
  year={2025},
  publisher={Nature Publishing Group UK London}
}

@article{dakheel2025optimizing,
  title={Optimizing smart grid load forecasting via a hybrid long short-term memory-XGBoost framework: Enhancing accuracy, robustness, and energy management},
  author={Dakheel, Falah and {\c{C}}evik, Mesut},
  journal={Energies},
  volume={18},
  number={11},
  pages={2842},
  year={2025},
  publisher={MDPI}
}

@article{javed2025sustainable,
  title={Sustainable energy management in the AI era: a comprehensive analysis of ML and DL approaches},
  author={Javed, Haseeb and Eid, Fatma and El-Sappagh, Shaker and Abuhmed, Tamer},
  journal={Computing},
  volume={107},
  number={6},
  pages={132},
  year={2025},
  publisher={Springer}
}

@article{tabaku2025utilizing,
  title={Utilizing artificial intelligence in energy management systems to improve carbon emission reduction and sustainability},
  author={Tabaku, Eda and Vyshka, Eli and Kap{\c{c}}iu, Rinela and Shehi, Alban and Smajli, Ensi},
  journal={Jurnal Ilmiah Ilmu Terapan Universitas Jambi},
  volume={9},
  number={1},
  pages={393--405},
  year={2025}
}

@article{boucetta2025wind,
  title={Wind power forecasting using a GRU attention model for efficient energy management systems},
  author={Boucetta, Lakhdar Nadjib and Amrane, Youssouf and Arezki, Saliha},
  journal={Electrical Engineering},
  volume={107},
  number={3},
  pages={2595--2620},
  year={2025},
  publisher={Springer}
}

@article{xu2026drl,
  title={DRL Energy Management for URT Considering Uncertainties of PV and Train Disturbance},
  author={Xu, Qian and Liu, Wei and Zhang, Xiaodong and Xia, Dingxin and Zhang, Jian and Tian, Zhongbei},
  journal={IEEE Transactions on Smart Grid},
  year={2026},
  publisher={IEEE}
}

@article{abishu2025multi,
  title={Multi-Agent DRL-Based Demand Response Optimization for IoT-Based Smart Home Energy Management Systems},
  author={Abishu, Hayla Nahom and Seid, Abegaz Mohammed and Erbad, Aiman and M{\'a}rquez-S{\'a}nchez, Sergio and Fernandez, Javier Hernandez and Corchado, Juan Manuel},
  journal={IEEE Internet of Things Journal},
  year={2025},
  publisher={IEEE}
}

@article{liu2025novel,
  title={Novel energy management strategy for fuel cell/battery hybrid energy systems combining MPC and deep reinforcement learning},
  author={Liu, Shengnan and Cheng, Hangyu and Jung, Seunghun and Kim, Young-Bae},
  journal={Energy Conversion and Management},
  volume={341},
  pages={120081},
  year={2025},
  publisher={Elsevier}
}

@article{wang2025safe,
  title={Safe deep reinforcement learning for building energy management},
  author={Wang, Xiangwei and Wang, Peng and Huang, Renke and Zhu, Xiuli and Arroyo, Javier and Li, Ning},
  journal={Applied Energy},
  volume={377},
  pages={124328},
  year={2025},
  publisher={Elsevier}
}

@article{song2025two,
  title={Two-layer deep reinforcement learning based port energy management strategy considering transportation-energy coupling characteristics},
  author={Song, Tiewei and Fu, Lijun and Zhong, Linlin and Fan, Yaxiang and Shang, Qianyi},
  journal={Energy},
  volume={318},
  pages={134659},
  year={2025},
  publisher={Elsevier}
}

\end{document}